\documentclass[aip,apl,showkeys,noeprint,superscriptaddress,preprint,longbibliography]{revtex4-1}
\usepackage{graphicx}
\usepackage{braket}
\usepackage{amsmath}
\usepackage{bm}  % for better bold face than amsmath
\usepackage[british]{babel}  % for better line-breaking
\usepackage[hidelinks,colorlinks,linkcolor=blue,citecolor=red,urlcolor=blue]{hyperref}
\usepackage[usenames, dvipsnames]{color}  % to have colored text
\usepackage{pdfcomment}  % to create comment in the pdf
\graphicspath{{../figs/}}

\begin{document}

\title{First-Principles Threshold Calculation of Photonic Crystal Surface-Emitting Lasers Using Rigorous Coupled Wave Analysis}

\author{Alex Y.\ Song}
\affiliation{Department of Electrical Engineering, Stanford University, Stanford, California 94305, USA}
\author{Akhil Raj Kumar Kalapala}
\affiliation{Department of Electrical Engineering, University of Texas at Arlington, Arlington, Texas 76019, USA}
\author{Weidong Zhou}
\affiliation{Department of Electrical Engineering, University of Texas at Arlington, Arlington, Texas 76019, USA}
\author{Shanhui Fan}
\email[]{shanhui@stanford.edu}
\affiliation{Department of Electrical Engineering, Stanford University, Stanford, California 94305, USA}

\date{\today}
\let\oldDelta\Delta
\renewcommand{\Delta}{\text{\scalebox{0.75}[1.0]{$\oldDelta$}}}

\begin{abstract}
We show that the threshold of a photonic crystal surface-emitting laser can be calculated from first-principles by the method of rigorous coupled wave analysis (RCWA), which has been widely used to simulate the response spectra of passive periodic structures. 
Here, the scattering matrix ($S$-matrix) of a surface-emitting laser structure with added gain is calculated on the complex frequency plane using RCWA, and the lasing threshold is determined by the value of gain for which the pole of the $S$-matrix reaches the real axis. 
This approach can be used for surface emitting laser structures in general, and is particularly useful for the surface emitting laser systems with complex in-plane structures. 

\end{abstract}


\maketitle

Surface-emitting lasers are advantageous over edge-emitting waveguide lasers in several aspects including
better beam shape and  the ease for integration as a two-dimensional array, and thus are widely used in optical communications and interconnects \cite{Chang-Hasnain2000,Koyama1989,Jewell1991,Soda1979,Lear1995,Martin-Regalado1997,Noda2017,Hirose2014,Matsubara2008,Kurosaka2010,Kosaka1999}.
The recent successful experimental demonstration of photonic-crystal surface emitting lasers (PCSELs) with high power, high beam quality, and beam-steering capability can further extend the usability of surface-emitting lasers in power-demanding applications such as free-space sensing \cite{Noda2017,Hirose2014,Chua2014,Chua2011,Matsubara2008,Kurosaka2010,Lu2008,Meier1999,Ryu2002,Vecchi2007,Vurgaftman2003,Notomi2001}.
Motivated by the experiments, there have been significant efforts in developing efficient simulation tools for PCSEL \cite{Wang2017,Peng2011,Liang2011,Liang2012,Yang2014,Sakai2010,Hung2012,Sakoda1999,Ryu2003}. 
Here, of particular interest is the capability to predict the threshold of PCSEL, taking into account the full complexity of the structure. 

In an edge-emitting waveguide laser, the threshold is typically calculated by equating the cavity round-trip gain to loss \cite{Chuang2009}.
However, in a PCSEL, the optical mode is defined by the 2D photonic crystal layer, and the cavity round-trip is not well defined.
Several recent works have developed coupled mode theory models for PCSEL \cite{Liang2011,Yang2014,Sakai2010,Wang2017}. These models typically treat the physics of PCSEL in terms of the coupling between small number of waveguide modes inside the photonic crystal layers. 
Such models provide significant insights into the operating mechanism of PCSEL. 
However, as a numerical method,  the coupled mode model makes uncontrolled approximations. 
For example, the use of only a small number of waveguide modes is difficult to justify in photonic crystal structures where index contrast can be quite large \cite{Skorobogatiy2003}. 
Also, these calculations typically obtain the transverse profile of the waveguide modes by considering a corresponding uniform dielectric waveguide, which again is approximate. This approximation in particular may influence the accuracy of the confinement factor which was used to compute the threshold in these analysis. 
\cite{Hung2012,Imada2002,Li2009,Maslov2004}. 

In the absence of gain, the PCSEL structure consists of multiple layers with periodic structures in some of the layers. Such a passive multilayer periodic structure can be readily treated using the rigorous coupled wave analysis (RCWA) method, for which several standard code packages are readily available \cite{Liu2012,Whittaker1999,Moharam1981,Tikhodeev2002}. 
In this paper, we show that these same RCWA code can be directly used, with very little modification, to compute the threshold of a PCSEL entirely from the first principle, taking into account the full complexity of the structure with no uncontrolled approximations. 
Conceptually, our development here builds upon the insights developed in the steady-state  \textit{ab initio} laser theory (SALT) \cite{Cerjan2014,Ge2010,Tureci2006}. It was shown in SALT that the threshold of a laser can be simulated in a linear calculation by adding gain to a passive structure, until for a specific gain value a pole of the scattering matrix ($S$-matrix) first crosses the real axis. Such a gain value then corresponds to the threshold gain. 
Previously, SALT has been applied in simulating non-regular laser cavities such as nano-disk lasers and random lasers \cite{Cerjan2014,Ge2010,Tureci2009}.
Here, we show that a combination of the concept of SALT with a numerical implementation in RCWA leads to a particularly convenient and powerful method for computing the threshold of PCSEL.

\begin{figure}[t]
    \includegraphics[width=244pt]{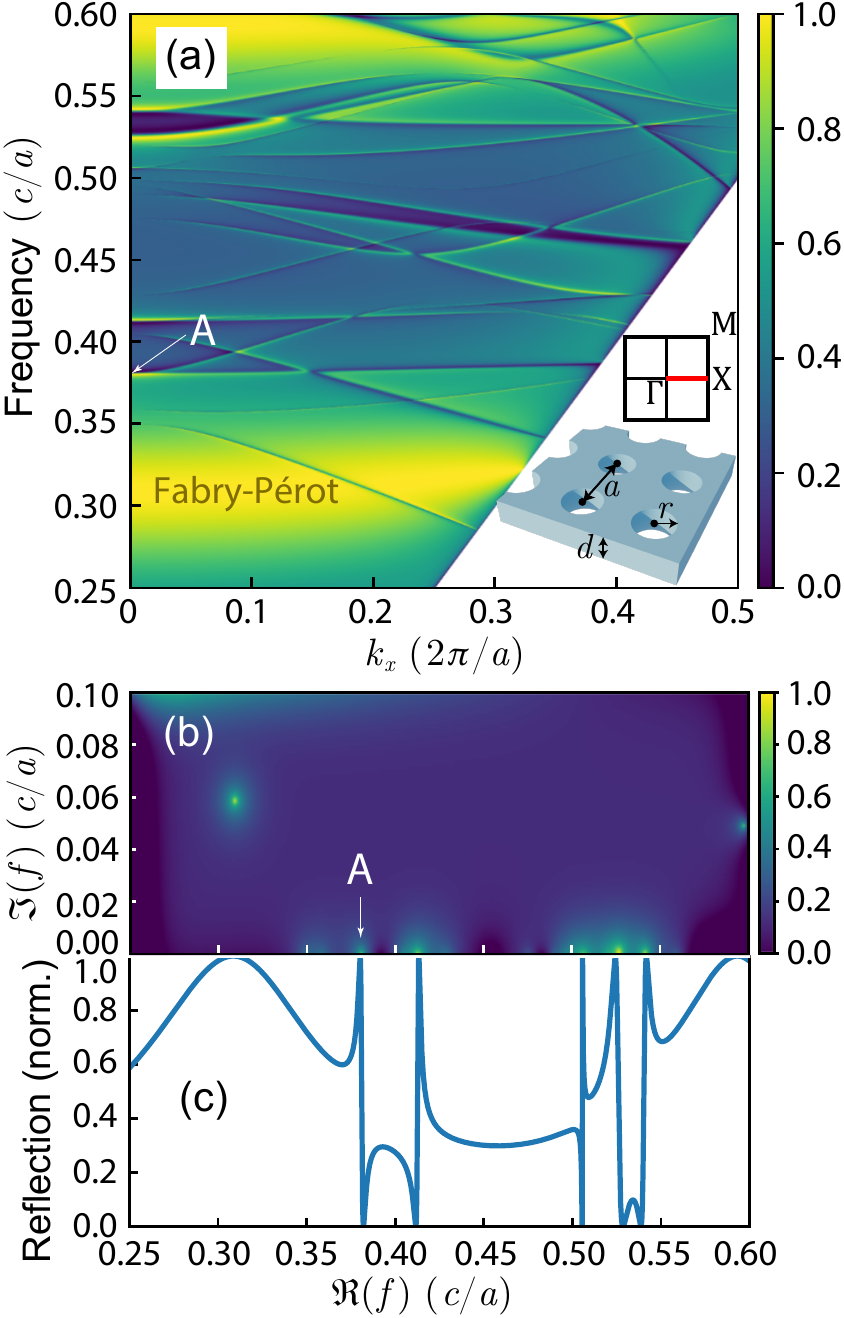}
    \caption{
    (a) Reflection spectra of a square lattice photonic crystal slab structure in the $\Gamma$-$X$ direction. Schematics of the photonic crystal slab and the first Brillouin zone are shown in the inset. $d$ is the slab thickness, $a$ is the lattice constant, and $r$ is the radius of the holes.
    % The doubly-degenerate mode with the lowest frequency at $0.38\,c/a$ is marked as mode A.
    (b) $\log(\det(S_{\Gamma}(\omega)))$ at $\Gamma$ point on the complex frequency plane. Bright points are the poles of the $S$-matrix.
    (c) The reflection spectra at $\Gamma$ point (normal incidence). The Fano resonances corresponds to the doubly degenerate modes at $\Gamma$.
    }
    \label{fig:PC_slab_bands}
\end{figure}

Surface-emitting lasers typically contain multiple layers with different refractive indices to confine light.
These layers can be either uniform or a 2D photonic crystal in the case of PCSEL. 
In a PCSEL, the photonic crystal slab layer is of critical importance since it defines the band structure and hence controls the lasing modal characteristics. Therefore, as a illustration of our method, 
here we first consider the calculation of  the threshold gain of a hypothetical laser structure consisting of a single 2D photonic crystal slab suspended in air. 
A schematic of the structure is shown in Fig.~\ref{fig:PC_slab_bands}. For this study, we assume the slab has a dielectric constant of $12$, representing that of a typical III-V semiconductor such as GaAs. 
Gain can be added as the imaginary part of the permittivity $\varepsilon_i$ in the slab. The holes and the surrounding vacuum has a dielectric constant of $1$. We assume the slab has a thickness of $d=0.5\,a$ where $a$ is the lattice constant. The holes have a radius of $r=0.2\,a$.

We start by simulating the passive structure in the absence of the gain using RCWA, which has been very widely used for this purpose. 
In Fig.~\ref{fig:PC_slab_bands}(a), we plot the intensity reflection coefficient as a function of both in-plane wavevector $k_x$ along the $x$-direction and frequency $f$. The in-plane wave vector varies along the $\Gamma$-$X$ direction in the first Brillouin zone of the photonic crystal. The reflection coefficients are calculated only for propagating modes that lies to the left of the light line $\omega = c_0 k_x$, where $c_0$ is the light speed in vacuum, and $\omega = 2\pi f$ is the angular frequency.
Also, in Fig.~\ref{fig:PC_slab_bands}(c) we plot the reflection spectrum at the $\Gamma$ point as a reference. 
% The slanted boundary on the right in Fig.~\ref{fig:PC_slab_bands}(a) represents the light-line.
In both Fig.~\ref{fig:PC_slab_bands}(a) and (c), we see the slow-varying features which corresponds to the Fabry-P\'{e}rot resonances of the structure, as well as the sharp spectral features that represent the guided resonances \cite{Fan2003,Fan2002}.
The plot in Fig.~\ref{fig:PC_slab_bands}(a), which shows the reflection spectra as a function of in-plane wavevector and frequency, thus in practice provides a simple way to visualize the photonic band structure of the guided resonance. 
At the $\Gamma$ point which corresponds to a plane wave normally incident upon the structure, due to the rotational symmetry, any bright mode must be two-fold degenerate. A bright mode is defined as a mode that can couple to externally incident plane wave. 
The lowest-frequency bright mode, with a frequency of approximately $0.38c/a$, is marked as mode A in Fig.~\ref{fig:PC_slab_bands}(a).

\begin{figure}
  \includegraphics[width=244pt]{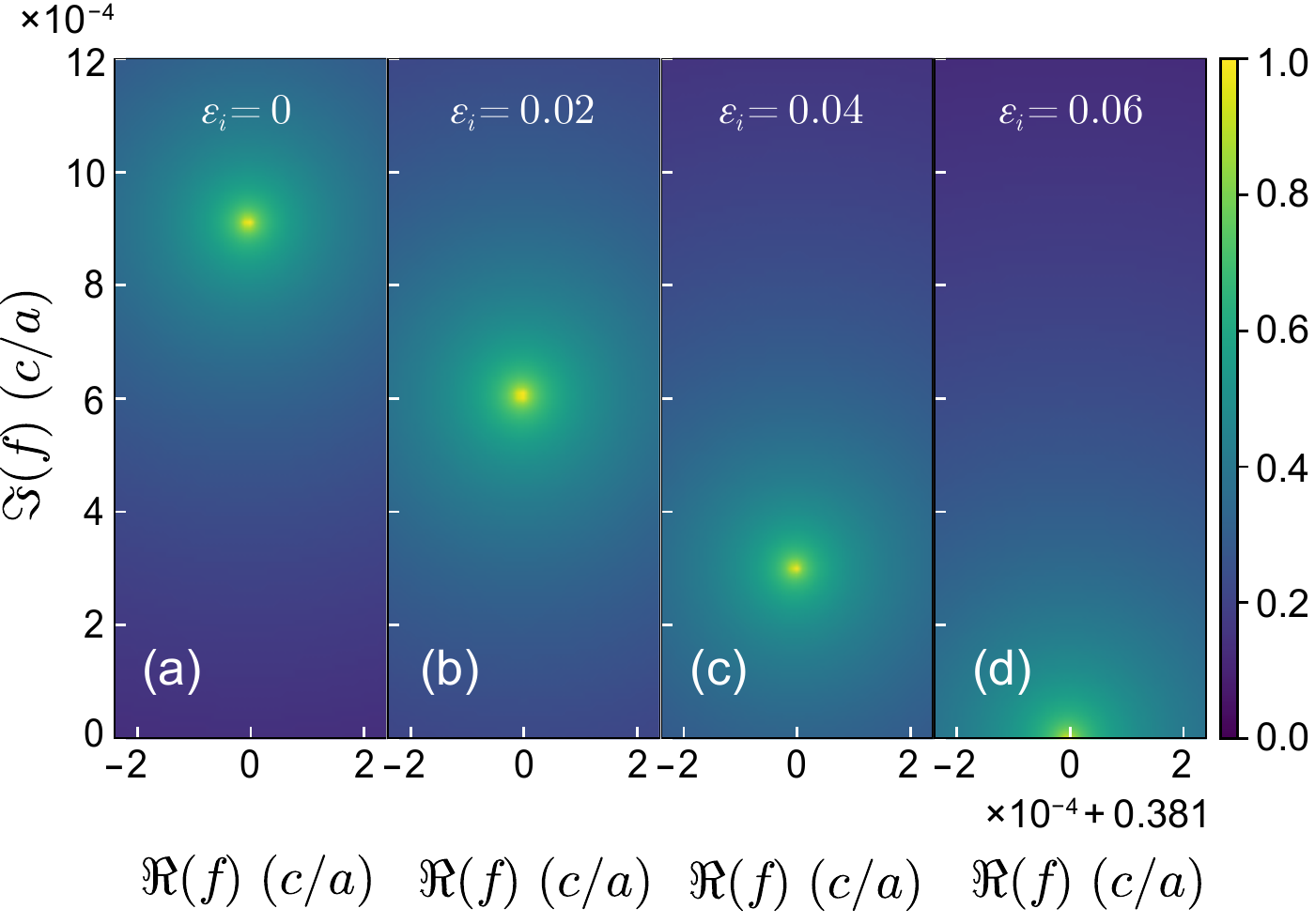}
  \caption{
  Threshold analysis of the photonic crystal slab in Fig.~\ref{fig:PC_slab_bands} incorporating gain.
  (a)-(d) Movement of the pole in the complex plane as the imaginary part of the dielectric constant in the photonic crystal slab $\varepsilon_i$ increases.
  }
  \label{fig:threshold}
\end{figure}

With RCWA, we can calculate the $S$-matrix of a PCSEL structure \cite{Liu2012,Whittaker1999,Moharam1981}, which relates the amplitudes of the input waves to those of the output waves, i.e.
\begin{equation}
    \label{eq:S_matrix}
    b = S_{\bm{k}}(\omega) a
\end{equation}
In Eq.~\ref{eq:S_matrix}, $\bm{k}$ is the Bloch wavevector defined in the first Brillouin zone of the crystal, it is conserved in the scattering process due to the in-plane periodicity of the structure. $a$ and $b$ are vectors, the components of which are the amplitudes of waves in the channels as labeled by the in-plane wave vectors $\bm{k}_n$ = $\bm{k} + n_1 \bm{G}_1 + n_2 \bm{G}_2$, where $n_1, n_2$ are integers and $\bm{G}_1, \bm{G}_2$ are the reciprocal lattice vectors. Since we assume vacuum outside of the slab, wave in each channel then has a wavevector component of $q_n = \sqrt{\omega^2/c^2-k_n^2}$ perpendicular to the slab. The channels are further labeled by whether the waves are above or below the slab, and by the polarization. 

In typical RCWA calculations, the frequency $\omega$ is assumed to be real. The channels can then be characterized as either propagating or evanescent in the direction perpendicular to the slab, depending on whether $q_n$ is real or imaginary. In our calculations, however, we will be interested in the analytic properties of $S_{\bm{k}}(\omega)$ in the complex $\omega$ plane \cite{Tikhodeev2002}. In this case, $q_n$ is in general complex for all $n$'s. With a complex or imaginary $q_n$, the incoming (outgoing) waves in a channel correspond to waves that spatially decays towards (away from) the slab. 

In the complex frequency plane, the frequency $\omega_p$, where $\det(S(\omega))$ diverges, define the pole of the $S$-matrix. A pole corresponds to a resonance of the structure. 
Due to causality, in a passive structure the imaginary part of the frequency of a pole must be non-negative. 
(Throughout the paper we follow the $e^{+i\omega t}$ convention for the complex field.)
In Fig.~\ref{fig:PC_slab_bands}(b) we plot $\log(\det(S_{\Gamma}(\omega)))$ on the complex frequency plane for the structure shown in Fig.~\ref{fig:PC_slab_bands}(a).
By comparing to the reflection spectra at $\Gamma$ shown in Fig.~\ref{fig:PC_slab_bands}(c), we can identify several types of resonances. 
The poles with the real part of $\omega_p$ at $0.31\,c/a$ and $0.59\,c/a$ corresponds to the Fabry-P\`{e}rot resonances of the slab.
The poles with their real part of $\omega_p$ in the range of $0.38$-$0.54\,c/a$, including mode A, are guided resonances. 
The imaginary parts of these guided-resonance poles are much smaller in magnitude as compared to the Fabry-P\`{e}rot poles, indicating that energy in these guided resonance leaks out of the structure at a much slower rate as compared to the Fabry-P\`{e}rot resonances. Finally, several poles, e.g. the one with a frequency of $0.35\,c/a$, is located directly on the real axis. These poles correspond to dark states in the band structure in Fig.~\ref{fig:PC_slab_bands}(a), in particular the singly degenerate modes at $\Gamma$, and they do not couple to free-space radiation \cite{Fan2002}. Thus they have no corresponding resonant features in the reflection spectra, as is seen by comparing Fig.~\ref{fig:PC_slab_bands}(b) and (c).

In the following, we proceed to compute the lasing threshold of mode A when gain is introduced into the system. In a PCSEL, lasing typically occurs at the band edge, where in-plane group velocity vanishes and hence the in-plane leakage rate is small. The band edge usually occur either at the center or the boundaries of the first Brillouin zone, unless especially designed \cite{Kurosaka2010}.
Moreover, in a semiconductor system, the gain spectrum of the material is relatively narrow. Thus, with proper choice of the geometric parameter to ensure that  the frequency of a particular band edge mode lies within the gain spectrum of the material, it is possible to design a PCSEL that selectively lases at a particular band edge mode. 

To compute the lasing threshold of mode A, we introduce a positive imaginary part to the permittivity and examine the position of the poles as the gain is increased. 
The enlarged plot of the pole of mode A is shown in Fig.~\ref{fig:threshold}(a).
With added gain in the structure, the net energy-loss rate in the mode is reduced.
Hence, the pole should move closer to the real axis as gain is increased.
The pole reaches the real axis under a $\varepsilon_i$ of $6\times 10^{-2}$. At this point the energy in mode A does not decay. The value of $\varepsilon_i$ therefore represents the lasing threshold of mode A.

\begin{figure}
    \includegraphics[width=245pt]{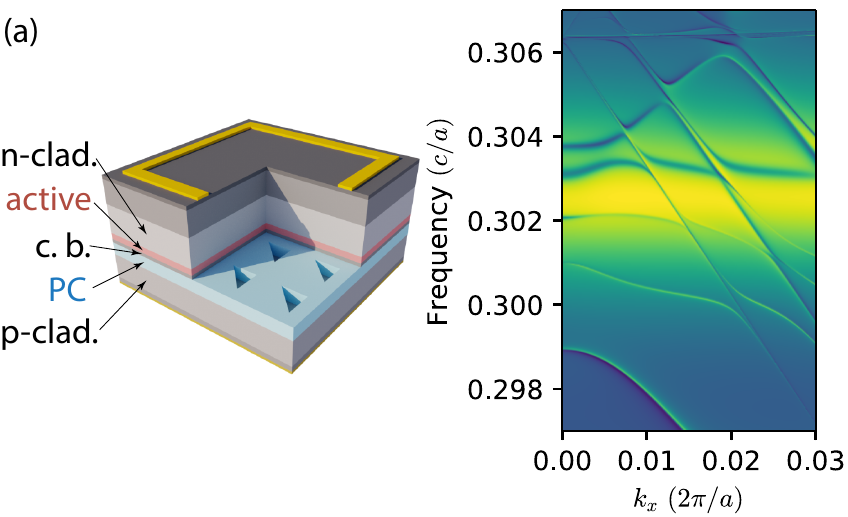}
    \caption{
    (a) Schematic of the PCSEL in Ref.~\onlinecite{Hirose2014}. The n-cladding layer, active layer, carrier blocking layer, photonic crystal layer, and the p-cladding layer are indicated in the plot.
    (b) Band structure of the PCSEL in $\Gamma$-$X$ direction. The modes A, B, C and D, following the notation in Ref.~\onlinecite{Hirose2014}, are identified and marked in the plot.
    }
    \label{fig:PCSEL_bands}
\end{figure}

The calculation approach discussed here can be readily applied to a realistic PCSEL structure. As an example, we study a PCSEL previously published in Ref.~\onlinecite{Hirose2014}. A schematic of the PCSEL is shown in Fig.~\ref{fig:PCSEL_bands}. 
For the optical computation, we only include the n-cladding layer, the active layer, the carrier blocking layer, the photonic crystal layer, and the p-cladding layer. 
The real part of the dielectric constant and the thickness of each layer are taken from Ref.~\onlinecite{Hirose2014}, and are reproduced in Table~\ref{table:PCSEL}. The lattice constant in the photonic crystal layer is $287\,$nm, and the triangular air holes have a side length of $175\,$nm. 
For simplicity, we assume the side wall of the air holes are vertical. More complex side-wall geometry can be incorporated in the RCWA calculation by dividing the photonic crystal layer into thinner layers with progressively changing hole sizes as an approximation. 

\begin{table}[t]
    \caption{Structure of the PCSEL in Ref.~\onlinecite{Hirose2014}.}
    \label{table:PCSEL}
    \begin{ruledtabular}
        \begin{tabular}{@{\hspace{0em}} c c c @{\hspace{0em}}}
        Layer  &   Dielectric Constant    &   Thickness (nm) \\
        \colrule
        n-cladding    &   9.747   &   2000 \\
        active    &   11.799   &   180 \\
        carrier blocking & 12.624 & 65 \\
        photonic crystal & GaAs: 12.624 / Air: 1.0 & 235 \\
        p-cladding & 10.713 & 1800 \\
        \end{tabular}
    \end{ruledtabular}
\end{table}

\begin{figure}
    \includegraphics[width=245pt]{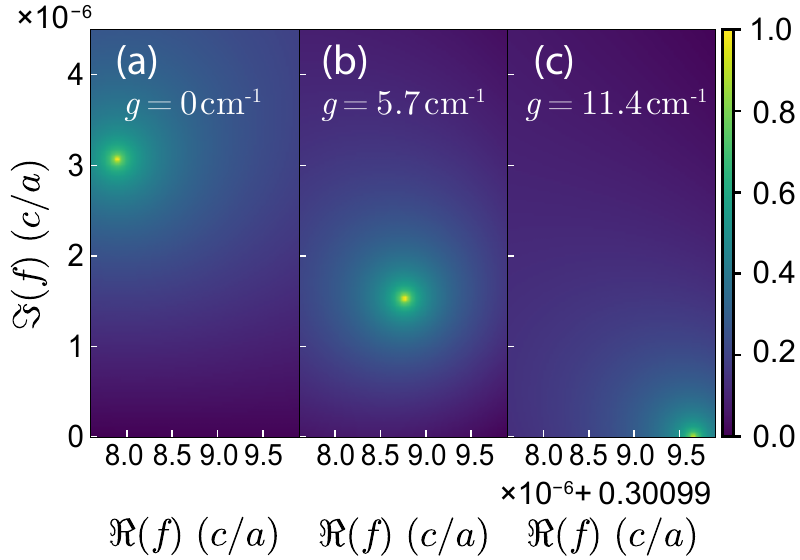}
    \caption{
    Threshold analysis of the PCSEL structure. (a)-(c) Movement of the pole corresponding to mode B, as the peak gain $g$ in the active layer is increased.
    }
    \label{fig:PCSEL_threshold}
\end{figure}

In Fig.~\ref{fig:PCSEL_bands}(b) we plot the reflection spectrum in the frequency range of $0.2975$-$0.3075\,c/a$.
The sharp spectral features here have the same shapes as the band structure measured in Ref.~\onlinecite{Hirose2014}. By comparing to the band structure to Fig.~4(a) of  Ref.~\onlinecite{Hirose2014}, we identify the lasing mode B as is marked in Fig.~\ref{fig:PCSEL_bands}(b).  The calculated band structure is shifted in frequency compared to Fig.~4(a) of  Ref.~\onlinecite{Hirose2014}. This may result from the deformations in size and shape of the air holes in the actual fabrication process. 
In Fig.~\ref{fig:PCSEL_threshold}(a) we show the computed $\det(S)$ as a function of real and imaginary parts of the complex frequency, in the vicinity of mode B, in the absence of gain. 
The $Q$ factor of the mode is calculated as $Q=\textrm{Re}(f)/\textrm{Im}(f)=9.8\times 10^4$, where $f$ is the complex frequency of the pole. 

To simulate the lasing threshold, we introduce optical gain to the active layer in the structure.
In reality, gain is provided by the optical transition between confined states in the quantum wells. Such gain always have dispersion. Here we assume a Lorentz model for the semiconductor gain, written as 
\begin{equation}
    \varepsilon(\omega) = \varepsilon_{b} + \frac{\Delta\varepsilon\omega_1^2}{\omega_1^2-\omega^2+i\Gamma\omega}
    \label{eq:Lorentz}
\end{equation}
In Eq.~\ref{eq:Lorentz}, $\varepsilon_{b}$ is the background permittivity, $\Delta\varepsilon$ is the oscillator strength, $\omega _1$ is the center angular frequency, and $\Gamma$ characterizes the gain bandwidth.
The peak gain $g$ is related to the oscillator strength by $g=\frac{1}{\Gamma c_0\sqrt{\varepsilon_{b}}}\Delta\varepsilon$.
Incorporating a frequency-dependent permittivity in RCWA is straightforward since RCWA is a frequency-domain method. 
Moreover, in order to treat such dispersion rigorously, we calculate the frequency-dependent permittivity on the complex plane by analytic continuation, i.e. using the complex $\omega$ in Eq.~\ref{eq:Lorentz}.
We vary the gain introduced into the structure by changing  the oscillator strength $\Delta\varepsilon$. 
For this study, we assume a nominal center wavelength of the gain profile at $940\,$nm, with a gain bandwidth of $5\%\,\omega_1$.
The background permittivity is $\varepsilon_b=11.799$.
As is shown in Fig.~\ref{fig:PCSEL_threshold},  the pole of the $S$-matrix moves towards the real axis as gain is increased, similar to the case of the photonic crystal slab. With a linear fitting, we retrieve the threshold condition as a peak gain of $g_{th}=11.4\,$cm$^{-1}$. 
This number is comparable, but somewhat smaller than that reported in Ref.~\onlinecite{Hirose2014}. This may be partially due to the fact that we have assumed uniform gain in the entire active layer, where as Ref.~\onlinecite{Hirose2014} assumed a quantum well gain medium. Since the electron wave functions are typically well confined in the quantum wells (QWs), optical gain should only benefit from the QWs, which is a fraction of the total thickness of the active layer. By comparing Fig.~\ref{fig:PCSEL_threshold}(a)-(c), we find that the real part of the frequency of the mode shifts as gain is increased. This is due to the change of the real part of the permittivity in the dielectric layer as gain is increased.

In the laser threshold simulation above, we have considered the intrinsic radiation loss, and neglected the doping induced waveguide loss in the layers. The latter can be straightforwardly incorporated as the imaginary parts of the dielectric constants in the corresponding layers.

In summary, we have shown that the threshold of a surface-emitting laser can be calculated from first-principles using RCWA. This method models the full 3D structure, and calculate the $S$-matrix of the structure on the complex frequency plane. In an active structure with gain, the threshold condition is obtained as the pole of the $S$-matrix reaches the real axis. 
This approach can be used for surface emitting laser structures in general. It is particularly useful for PCSEL which have complex periodic in-plane structures, where the conventional approach of threshold analysis in waveguide lasers does not apply.

\begin{acknowledgments}
This work is supported in part by the Department of Defense Joint Technology Office under Grant No. N00014-17-1-2557. 
\end{acknowledgments}

% \bibliography{library.bib}

%Control: production of eprint (-1) disabled
%

\end{document}